\documentclass[aps,onecolumn,prd,showpacs,showkeys,preprintnumbers,superscriptaddress,nobibnotes,notitlepage,floatfix,longbibliography,nofootinbib]{revtex4-2}

\pdfoutput=1

\usepackage{graphicx} 
\usepackage{mathtools}
\usepackage{enumitem}
\usepackage{amssymb,amsmath}
\usepackage{float,xcolor}

\def\be{\begin{equation}}
\def\ee{\end{equation}}
\def\bea{\begin{eqnarray}}
\def\eea{\end{eqnarray}}

\def\pc{{\rm pc}}
\def\kpc{{\rm kpc}}

\begin{document}

\title{Dark Matter Velocity Distributions: Comparing Numerical Simulations to
Analytic Results}

\author{Katharena Christy}
\thanks{{\scriptsize Email}: \href{mailto:jkumar@hawaii.edu}{chri3448@hawaii.edu}}
\affiliation{Department of Physics and Astronomy, University of Hawai'i, Honolulu, HI 96822, USA}

\author{Jason Kumar}
\thanks{{\scriptsize Email}: \href{mailto:jkumar@hawaii.edu}{jkumar@hawaii.edu}}
\affiliation{Department of Physics and Astronomy, University of Hawai'i, Honolulu, HI 96822, USA}

\author{Louis E.~Strigari}
\thanks{{\scriptsize Email}: \href{mailto:jkumar@hawaii.edu}{strigari@tamu.edu}}
\affiliation{Department of Physics and Astronomy,
Mitchell Institute for Fundamental Physics and Astronomy,
Texas A\&M University, College Station, TX  77843, USA}

\begin{abstract}
We test the consistency of  dark
matter velocity distributions obtained from dark matter-only numerical simulations with analytic predictions,
using the publicly available Via Lactea 2 dataset as an example.
We find that, well inside the scale radius, the velocity distribution obtained from numerical
simulation is consistent with a function of a single integral of motion -- the energy --
and moreover is consistent with the result obtained from Eddington inversion.  This indicates that
the assumptions underlying the analytic result, namely, spherical symmetry, isotropy, and a
static potential, are sufficiently accurate to govern the coarse properties of the velocity
distribution in the inner regions of the halo.  We discuss implications for the behavior of
the high-velocity tail of the distribution,
which can dominate dark matter annihilation from a $p$- or $d$-wave state.
\end{abstract}

\maketitle

\section{Introduction}

The velocity distribution of particle dark matter (DM) in (sub)halos is an important input to
dark matter direct and indirect detection searches.  There are two main strategies for
studying this distribution: large $N$ numerical simulations~\cite{Diemand:2006ik,Vogelsberger:2007ny,Diemand:2008in,Stadel:2008pn},
and analytic analyses of the
phase space distribution~\cite{Widrow:2000dm,Evans:2005tn}.  The advantage of simulations is that they do not require the use
of approximations, such as spherical symmetry or isotropy,\footnote{Although isotropy is assumed in
analytic analyses using Eddington inversion, which we consider, there are other approaches which do not require
this assumption (see, for example,~\cite{Wojtak:2008mg,Mamon:2012yb}).  An axisymmetric approach has been considered
in Ref.~\cite{Petac:2021cjm}, for example.} and incorporate the details of the merger
history.  The advantage of analytic analyses,
apart from computational simplicity, is that they allow one to learn broad lessons which apply
beyond the details of an individual simulation.  Our goal in this work is to investigate the
extent to which the results of numerical simulations of the dark matter velocity distribution
match the general predictions of analytic analyses.

The answer to this question has significant implications for dark matter detection strategies.
The connection between the dark matter velocity distribution and direct detection has been
well studied~\cite{Fox:2010bu}.  For many dark matter candidates, it is only the high speed tail of the DM
distribution which can deposit enough energy in the detector to produce a recoil above threshold.
Moreover, the velocity distribution affects the annual modulation of a direct detection signal~\cite{Freese:2012xd}.
The connection between the DM velocity distribution and indirect detection has been less well studied
because, if the DM annihilation cross section is velocity-independent (the most commonly studied scenario),
then the annihilation rate will depend only on the density distribution, not the velocity distribution.
But there are a variety of particle physics models in which the annihilation cross section is velocity-dependent,
and in these cases, the astrophysical $J$-factor which controls the annihilation rate in any particular astrophysical
target will depend in detail on the velocity distribution~\cite{2009PhRvD..79h3525R,2014GrCo...20...47B,Ferrer:2013cla,Boddy:2017vpe}.

For example, for models of $p$- or $d$-wave dark matter annihilation, the dark matter indirect detection signal
receives an enhanced contribution from the annihilation of particles in the high-speed tail of
the velocity distribution.
In numerical simulations, one finds that in any radial shell, in simulations with baryons, the
velocity distribution is well fit by a Maxwell-Boltzmann distribution with an exponential tail
(see, for example,~\cite{Piccirillo:2022qet,Blanchette:2022hir}). For dark matter-only simulations, the Maxwell-Boltzmann
distribution fit is still reasonable but not as robust~\cite{Kuhlen:2009vh,BeraldoeSilva:2013efk}.  More generally,
given the large uncertainties in these fits, it is difficult to quantify how different the behavior of the high-speed tail
of DM-only simulations is from the Maxwell-Boltzmann distribution.
On the other hand, analytic analyses suggest that, under certain assumptions, the high-speed tail falls
off only as a power law near the center of the halo (see, for example,~\cite{Boucher:2021mii,Kiriu:2022bjq}).
Although the broad features of these distributions
are similar, the differences are significant.  If the velocity distribution only falls off as a power-law,
then the power-law enhancement of the cross section in the case of $p$-/$d$-wave annihilation can compensate,
causing the indirect detection signal to be dominated by the small fraction of particles in the high-speed
tail.  This tail is much less significant if the velocity distribution exhibits an exponential falloff at
high speed.  It is thus important to know if the results of numerical simulations, though well fit
by a Maxwell-Boltzmann distribution, are generally consistent with the predictions of analytic analyses.
Particularly since the high-speed tail is necessarily relatively poorly sampled in numerical simulations, it is helpful
to know if analytic results for the asymptotic behavior of the tail can be trusted.
More generally, if the results of numerical simulations are consistent with those of analytic analyses, then
credence is lent to the idea that the approximations which underlie the analytic analyses are sufficiently
good.

In this work, we consider the publicly available results of the Via Lactea 2 (VL-2) DM-only numerical
simulation~\cite{Diemand:2008in},
consisting of
$10^5$ particles randomly drawn from the $10^9$ particles in their sample.  We find that, for radial
distances less than half the scale radius, the velocity distribution
inferred from these particles is broadly consistent with being a function of a single integral of motion -- the energy.
This is the result predicted from
analytic analyses, under the assumptions of spherical symmetry, isotropy, and time-invariance.
Although these assumptions are not exact, deviations from these assumptions are apparently small enough
that they do not have a large effect on coarse features of the velocity distribution in the inner regions of the
halo.  More specifically, we find that the velocity distribution obtained from VL-2 is consistent with the
result obtained from Eddington inversion.
The consistency between Eddington inversion and the results of numerical simulations was also
discussed, though more qualitatively, in Ref.~\cite{Lacroix:2020lhn}.  We compare our approach and our results with
theirs.

The plan of this paper is as follows.  In Section~\ref{sec:analytic}, we will review results from
an analytic analysis of the phase distribution.  We compare these to publicly-available results obtained from the VL-2
numerical simulation in Section~\ref{sec:numerical}.  We conclude in Section~\ref{sec:conclusion}.

\section{Analytic Analysis of the Phase Space Distribution}
\label{sec:analytic}

This discussion follows the results of several standard texts
(see, for example,~\cite{goldstein:mechanics,2008gady.bookB}).
In general, the velocity distribution $f (\vec{r}, \vec{v}, t)$ is a function of seven variables.
We begin with four initial assumptions:
\begin{enumerate}[label=(\roman*)]
\item{the DM velocity distribution (as well as the distribution of any baryonic matter) is spherically symmetric,}
\item{the DM velocity distribution (as well as the distribution of any baryonic matter)  is independent of time, }
\item{each DM particle is subject only to central forces which depend only on its radial position, }
\item{ and the DM velocity distribution is isotropic (optional).}
\end{enumerate}
If the only relevant forces are gravitational, then assumptions
(i) and (ii) together imply assumption (iii).  Although this is the standard scenario, in the interests of generality,
we will retain (iii) as an independent assumption.  Of course, none of these assumptions will be exactly
true.
We emphasize that numerical simulations, including VL-2 (see, for example,~\cite{Zemp:2009ff}), exhibit significant
deviations from these
assumptions.
Our goal will be to examine the consequences of these assumptions, and compare the
resulting predictions to results from numerical simulations, in which none of these assumptions are
made.

Assumptions (i)-(iii) imply that the velocity distribution is only a function
of three variables: $r$, $v_r$ and $v_\perp$, where $r = |\vec{r}|$, $v_r = \vec{v} \cdot \hat r$,
and $v_\perp = \sqrt{\vec{v}^2 - v_r^2}$.  Moreover, these
three assumptions imply that each dark matter particle may be thought of as moving under the influence
of a single time-invariant central potential.  $\vec{r}$ and $\vec{v}$ are thus functions of time and
six integrals of motion.

Two integrals of motion determine the plane of motion, and another determines the
orientation of the orbit in the plane.  The remaining three are the energy, the magnitude of the
angular momentum, and a constant $t_0$ which identifies the time zero-point.  Spherical
symmetry implies that $f$ is independent of the first three integrals of motion, and thus is a function of
only three variables, as expected.  Liouville's theorem
further implies that the phase space density is invariant under time-translation, so
$f$ is also independent of $t_0$.  Thus, assumptions (i)-(iii) imply that $f$ is a function
only of the energy and the magnitude of the angular momentum.
If assumption (iv) also holds, then $f$ depends only on $r$ and $v = |\vec{v}|$, or equivalently,
$f$ is a function of energy alone.

Given assumptions (i)-(iii), the density distribution is a function only of $r$, given by
\bea
\rho (r) &=& \int_0^{v_{esc}(r)} d^2 v_\perp dv_r~ f(r, v_r, v_\perp) ,
\nonumber\\
&=& 2\sqrt{2}\pi
\int_0^{\sqrt{2}r \sqrt{\Phi(\infty)- \Phi(r)}} dL~
\int_{L^2/2r^2 + \Phi(r)}^{\Phi(\infty)} dE ~ \frac{L}{r^2}\frac{1}{\sqrt{E - \frac{L^2}{2r^2}-\Phi(r)}}  f(E,L) ,
\label{eq:rho_and_fEL}
\eea
where $E$ and $L$ are the energy and angular momentum per unit mass, respectively.  $\Phi$ is the gravitational
potential, and
\bea
E &=& \frac{1}{2} \left( v_r^2 + v_\perp^2 \right) + \Phi (r),
\nonumber\\
L &=& r v_\perp ,
\nonumber\\
v_{esc}(r) &=& \sqrt{2 \left( \Phi (\infty) - \Phi (r) \right) } .
\eea
If $f$ is independent of $L$, then one may perform the integral over $L$, yielding
\bea
\rho (r) &=& 4\sqrt{2} \pi \int_{\Phi(r)}^{\Phi(\infty)} dE ~\sqrt{E - \Phi(r)}~ f(E) .
\label{eq:rho_and_fE}
\eea

We can use these formulae to test if the velocity distribution is determined by these two integrals of motion.
If assumptions (i)-(iv) hold, and $f$ depends only on $E$, then in any thin radial shell of volume $dV$,
the number of particles within the energy range $[E,E+dE]$ is given by
\bea
N (E,r) &=&  \left[4\sqrt{2} \pi \sqrt{E - \Phi(r)}~dE~dV \right] f(E) .
\label{eq:N_and_fE}
\eea
Scaling out the factor in brackets, one can use the particle count per energy bin in a numerical simulation
to determine $f(E)$ for each radial bin.\footnote{Assuming spherical symmetry and that all forces are
gravitational,
$\Phi (r) = G_N \int_0^r dx~M(x)/x^2$, where $M(r)$ is the mass of the particles enclosed within
radius $r$, and we have set $\Phi(0)=0$.}  If $f$ only depends on $E$, then these functions will be identical,
differing only in the range of energies (from $\Phi(r)$ to $\Phi(\infty)$) which are sampled in each radial shell.

If only assumptions (i)-(iii) are good approximations, then
$f$ will depend on $E$ and $L$.  The integral over $L$ is then non-trivial, and  the rescaling
above would not remove all radial dependence.  As a result, the functions $f(E)$ found in different radial bins would
not agree.  In that case, one would  use the particle count in radial bins and in bins of $E$ and $L$ in
order to determine $f (E,L)$, after rescaling by the appropriate function of $E$, $L$ and $\Phi(r)$ found in
eq.~\ref{eq:rho_and_fEL}.  But if even assumptions (i)-(iii) are not good approximations, then $f$ would
not be expected to depend only on $E$ and $L$.  In this case, even the functions $f(E,L)$ obtained from the
particles counts in different radial bins would not agree.

In particular,  if we derive the velocity distribution from numerical simulation data in two
non-overlapping radial bins, then these velocity distributions have support over
disjoint regions of phase space which are not related by spherical symmetry,
and are a priori independent.  What relates them is the fact that, given our assumptions,
the integrals of motion $E$ and $L$ remain constant as a particle moves from one region of phase
space to another.  This need not be true if assumptions (i)-(iii) do not hold, as energy or
angular momentum could be transferred from one particle to another.

\section{Comparison to Numerical Simulation}
\label{sec:numerical}

We consider the Via Lactea 2 DM-only simulation~\cite{Diemand:2008in}, consisting of $10^9$ particles.  If fit to
a generalized Navarro-Frenk-White (gNFW) profile of the form
$\rho (r)  = \rho_s / [(r/r_s)^\gamma (1 + r/r_s)^{3-\gamma}]$~\cite{Zhao:1996mr}, the best fit value for the
inner slope of the halo is $\gamma = 1.24$, with a scale radius of $r_s = 28.1~\kpc$
and a scale density $\rho_s = 0.0035 M_\odot / \pc^3$.
The radius of convergence is $0.38~\kpc$, and is not expected to affect our results.
We use the $10^5$ randomly selected particles which
have been made publicly available.  Fitting the enclosed mass of this subset of particles as a function
of radius to a gNFW profile with $\gamma = 1.24$ yields $\chi^2 /dof = 0.56$ (using 9 radial bins extending
between $0.05r_s$ and $0.5r_s$).
This indicates that the publicly available points
do indeed constitute a representative sample which may be used to study the velocity distribution of
this halo.

For simplicity, we use the dimensionless variable $\tilde r \equiv r / r_s$.
We focus on the region $\tilde r < 0.5$ (containing $\sim 3600$ particles), which lies well within the scale radius, and
for which we expect deviations from spherical symmetry and isotropy to be relatively small~\cite{Zemp:2009ff}.
Understanding this region should give us a good understanding of the dark matter velocity
distribution in the region of phase space most relevant for indirect detection, as dark matter annihilation
is concentrated in the innermost regions of the halo, especially
if the inner slope is relatively steep.

We divide the range $\tilde r < 0.5$ into 5 radial shells (A-E), and in each, we estimate
deviations from spherical symmetry and isotropy.  In particular, we expand the density distribution
in spherical harmonics and compute the associated expansion coefficients $a_{\ell m}$ for $\ell \leq 1$.
We use the definition
$a_{\ell m} \equiv [\int dV \rho (r, \theta, \phi) Y_{\ell m} (\theta, \phi)]/[\int dV \rho (r, \theta, \phi)]$,
where both integrals are taken over the volume of the radial bin.
We also compute the anisotropy coefficient $\beta \equiv 1 - [\langle v_\perp^2 \rangle / 2 \langle v_r^2 \rangle]$.
We define the radial extent of each region, and present the $a_{\ell m}$ and $\beta$ (with
uncertainty)\footnote{Adopting standard propagation of errors, we use
$\delta \beta = \frac{\langle v_\perp^2 \rangle}{2 \langle v_r^2 \rangle}
\left[\frac{\langle v_r^4 \rangle - \langle v_r^2 \rangle^2}{N\langle v_r^2 \rangle^2}
+ \frac{\langle v_\perp^4 \rangle - \langle v_\perp^2 \rangle^2}{N\langle v_\perp^2 \rangle^2}  \right]^{1/2}$, where
$N$ is the number of particles in the radial bin.  }
in Table~\ref{table:AandBeta}.
In each region, the anisotropy parameter $\beta$ and the dipole terms in the density distribution are relatively small,
though non-zero.

Note, however, that $\beta$ is really a spherically-averaged measure of anisotropy.  One generally finds, even
well within the cusp,
that velocity ellipsoids tend to align with the major axis of the halo~\cite{Wojtak:2013eia} in many numerical simulations,
including VL-2~\cite{Zemp:2008gw}.  As a result, despite the relatively small value of $\beta$, the assumption of isotropy
is at best a coarse-grained approximation.

We expect that the assumption of a static distribution should be reasonable, at a coarse-grained
level, because the relaxation time is typically large when considering a self-gravitating system.  We can test this
hypothesis by considering the parameter $q = 2 (\sum E_{kin}) / (- \sum F \cdot r)$, where the sum is taken over
all particles satisfying $r  \leq r_{max} = 24 r_s$.  Note, the choice of $r_{max}$ is dictated by the fact that,
for $r > r_{max}$, the enclosed mass no longer is a good fit to the gNFW profile, indicating that one has reached
the edge of the halo.  For a virialized system in equilibrium, one should find $q=1$.  Instead we find
$q = 1.09$, quantifying the extent to which there are deviations from equilibrium.

\begin{table}[!h]
\begin{tabular}{|c|c|c|c|c|c|}
  \hline
  region & $a_{00}$ & $a_{10}$ & Re $a_{11}$ & Im $a_{11}$ & $\beta$ \\
   \hline
  $0 < \tilde r < 0.1$  (A) & $0.28$ & $0.0028$  & $0.0095$ & $0.0000198$   & $ 0.04 \pm 0.12$ \\
   $0.1 < \tilde r < 0.2$ (B) & $ 0.28$ & $ 0.01$  &  $ 0.01$ & $ 0.0024$   & $0.03 \pm 0.08$ \\
   $0.2 < \tilde r < 0.3$ (C) & $0.28$ & $0.01$ & $0.007$ & $0.027$ & $0.14 \pm 0.05$ \\
    $0.3 < \tilde r < 0.4$ (D) & $0.28$ & $-0.005$ & $0.008$ & $0.02$ & $0.16 \pm 0.05$ \\
 $0.4 < \tilde r < 0.5$ (E) & $0.28$ & $0.001$ & $-0.015$ & $0.022$ & $0.13 \pm 0.04$ \\
  \hline
\end{tabular}
\caption{The radial extent of each of the five regions considered, along with the $a_{\ell m}$ ($\ell \leq 1$) and $\beta$.}
\label{table:AandBeta}
\end{table}

We now investigate the extent to which the velocity distribution in the range $\tilde r < 0.5$ can be well-represented as a function
of energy alone.  We divide each region into two subregions (denoted ``1" (inner) and ``2" (outer)) at the radial midpoint.\footnote{For
example, subregions A1 and A2 extend from $0 < \tilde r < 0.05$ and $0.05 < \tilde r < 0.1$, respectively.}
Within each region, we choose a common
energy binning for both subregions.  In each subregion, we then determine the velocity distribution $f_i$ in the $i$th energy bin using the relation
$f_i = N_i / [4\sqrt{2} \pi \sqrt{E - \Phi (\tilde r)} \Delta E \Delta V]$, where $N_i$ is the number of particles in that energy bin,
$\Delta E$ is the width of the energy bin, and $\Delta V$ is the spatial volume of the subregion.  Here, $E$ is evaluated at the midpoint of
the energy bin, and $\Phi (\tilde r)$ is taken to be the average value of the potential over all particles in the subregion, assuming that the density
distribution follows a gNFW form with $\gamma = 1.24$.  Note, we ignore energy bins for which $E_{avg} - \Phi_{avg} (\tilde r) <0$.
More generally, because the scaling factor in eq.~\ref{eq:N_and_fE} is averaged over the energy and spatial bin, there will be associated binning
error in the resulting velocity distribution.

We then compare the values of $f(E)$ in the two subregions, to see if they are statistically consistent.
The standard deviation of $f_i (E_j)$ in the $i$th radial bin and $j$th energy bin is taken to be
$\sigma_{ij} = f_i(E_j) / \sqrt{N_{ij}}$, where $N_{ij}$ is the number of particles in the $i$th radial bin and
$j$th energy bin.  We compare the values of $f(E)$ in
subregions 1 and 2 by computing $\chi^2 \equiv \sum_j [f_1 (E_j) - f_2 (E_j)]^2/[\sigma_{1j}^2 + \sigma_{2j}^2]$.
Our results for the 5 pairs of
subregions are presented in  Table~\ref{table:chi2}.  We generally find consistency, indicating that, to a good approximation,
the dependence of the velocity distribution on $r$ and $v$ arises from the dependence of $E$ on both.
Note that some particles within radial subregion A1 lie inside the radius of convergence ($\tilde r < 0.135$).
But due to the volume factor,  $< 10\%$ of particles in this bin lie inside the radius of convergence, so including these particles
does not affect our results significantly.

\begin{table}[h]
\begin{tabular}{|c|c|c|}
  \hline
    pair & \# of energy bins & $\chi^2/dof$  \\
   \hline
A1,A2 & $8$ & $0.92$ \\
B1,B2 & $10$ & $1.42$ \\
C1,C2 & $9$ & $0.71$ \\
D1,D2 & $8$ & $1.11$ \\
E1,E2 & $10$ & $0.95$ \\
  \hline
\end{tabular}
  \caption{For each region A through E, the number of energy bins used, and
  the $\chi^2 /dof$ between $f (E)$ of the two subregions.
  Note that the number of energy bins varies slightly from region to
  region, because energy bins with $E_{avg} - \Phi_{avg} (\tilde r) <0$ are rejected.}
  \label{table:chi2}
\end{table}

\begin{figure}[h]
\centering
\includegraphics[width=\textwidth]{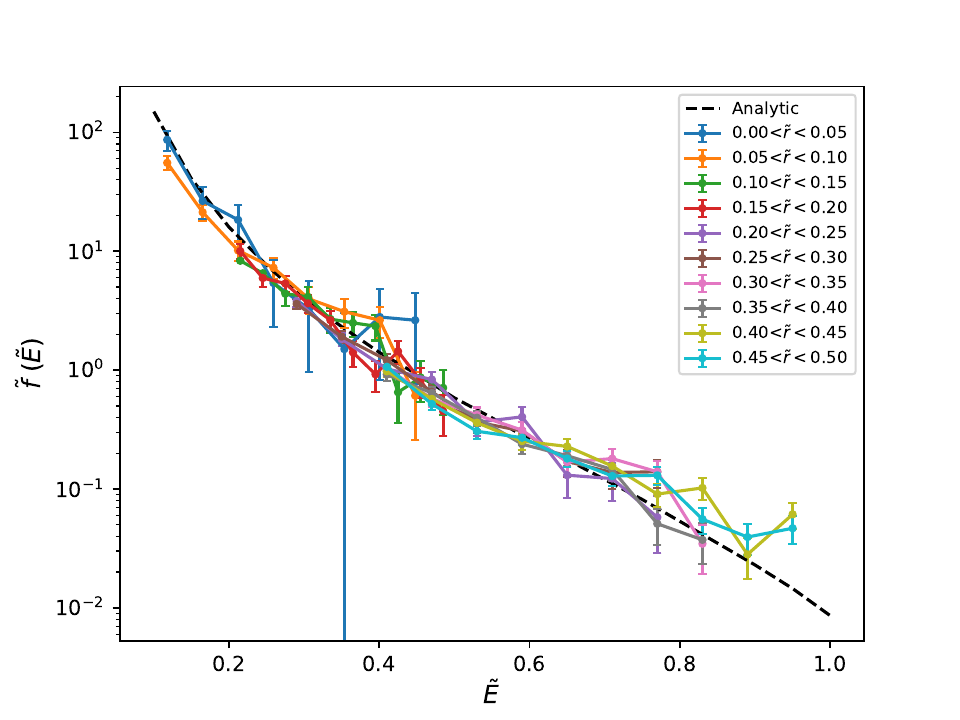}
\caption{$\tilde f (\tilde E)$ for ten radial bins from 0 to $0.5 \tilde r_s$, with each bin
of width $0.05 \tilde r_s$, as labelled.  Also plotted (dashed) is the analytic result obtained from
Eddington inversion~\cite{Boucher:2021mii}, assuming $\gamma = 1.25$.}
\label{fig:FE_main}
\end{figure}

Finally, we can compare the $f(E)$ found in these ten subregions to the analytic expression
one would obtain from Eddington inversion assuming a gNFW profile with inner slope of $\gamma = 1.24$.
Essentially, Eddington inversion amounts to inverting eq.~\ref{eq:rho_and_fE} with an inverse Abel integral
transform, allowing one to numerically solve for $f(E)$, given an ansatz for $\rho (r)$:\footnote{Note that,
in general, the result obtained from Eddington inversion depends on how the halo truncates, for example, at the tidal radius.
We consider Eddington inversion for an infinite gNFW profile, for which the surface term is irrelevant.  This is expected to provide a
good approximation to the velocity distribution well inside the cusp, since only a very small fraction of particles
deep within the cusp will be energetic enough to reach the edge of the halo.}
\bea
f (E) &=& \frac{1}{\sqrt{8} \pi^2} \int_{E}^{\Phi (\infty)}
\frac{d^2 \rho}{d\Phi^2} \frac{d\Phi}{\sqrt{\Phi - E}} .
\label{eq:Eddington}
\eea
For simplicity, we define the constant $E_0 = 4\pi G_N \rho_s r_s^2$, and the dimensionless quantities
$\tilde E = E/E_0$, $\tilde f = f / (\rho_s E_0^{-3/2})$.
We plot $\tilde f( \tilde E)$  in the ten subregions,
along with the analytic result (obtained using Eddington inversion for the
case of a gNFW profile with $\gamma = 1.25$~\cite{Boucher:2021mii}), in Figure~\ref{fig:FE_main}.
The velocity distribution obtained from VL-2 data is consistent
with the result obtained from Eddington inversion.  Given this consistency, there is little to
be gained from attempting to determine the dependence of the velocity distribution on angular
momentum, given this limited data set.

We can compare these results to those of Ref.~\cite{Lacroix:2020lhn}, which compared
the results of three (Milky Way-sized) numerical simulations to the results of Eddington inversion, for both DM-only scenarios
and for runs including the effects of baryons.  Ref.~\cite{Lacroix:2020lhn} also found general agreement between
Eddington inversion and the results of their numerical simulations, for both DM-only runs and runs including
baryonic effects.  In particular, Ref.~\cite{Lacroix:2020lhn} found that Eddington inversion generally provided
a better fit to numerical simulation data than a Maxwell-Boltzmann model.  However, the consistency of numerical
simulation data with a velocity distribution dependent only on $E$ is only presented qualitatively in Ref.~\cite{Lacroix:2020lhn}.  In particular, $f(E)$ is derived from radial bins which extend from well inside
the cusp to $10^4~\kpc$.  At such large radii, the halo has significant deviations from isotropy.  As a result,
for fixed $E$, the values of $f(E)$ found in Ref.~\cite{Lacroix:2020lhn} span more than 2 orders of magnitude.  As no
statistical uncertainties arising from numerical sampling are provided, Ref.~\cite{Lacroix:2020lhn} can only
provide a qualitative statement of agreement between numerical simulation and Eddington inversion.

Ref.~\cite{Lacroix:2020lhn} does account for statistical uncertainty due to numerical simulation when comparing
the velocity distributions obtained from 4 radial bins to the predictions of Eddington inversion.  But again, they
can only find qualitative agreement in these four radial bins.  Indeed, they find $\chi^2/dof$ significantly larger
than 1.  This may be related to the fact that two of four radial bins they considered had radii comparable to or larger
than the scale radius; if anisotropy is important for those bins, then the velocity distribution would not be expected
to match the results of Eddington inversion.  Moreover, the Eddington inversion result itself was derived from a fit
to the density profile of the entire halo, including regions at large $r$ for which deviations from spherical symmetry
and equilibrium could be important.

By contrast, our results show much more quantitative agreement, not only between the velocity distribution
derived from VL-2 data and that derived from Eddington inversion, but also between the velocity distribution derived
from VL-2 data in different radial bins with each other.  This may be due to the fact that our method focused on fitting
the density and gravitational potential in the innermost regions of the cusp, for which the approximation of a velocity
distribution which depends only on $E$ is expected to be better.  A more detailed comparison of these results would be
an interesting topic of future work.

In this context, we note also that Eddington inversion seems to yield a slight but systematic
overestimate of the velocity distribution for $0.4 < \tilde E < 0.6$ and a similar underestimate of the
velocity distribution for $0.7 < \tilde E < 1$.  In particular, for these energies, the velocity distributions obtained
from different VL-2 radial bins are more consistent with each other than they are with the result of Eddington inversion.
Of course, if the velocity distribution is truly a function of energy alone, then it should be exactly equal to the
Eddington inversion result.  But we have only focused on the region of small energy, or equivalently, well within the
cusp.  There are good reasons to believe that, even if the velocity distribution is well described as a function of
$E$ well within the cusp, it might depend non-trivially on $L$ at large distances.  In this case, Eddington inversion
may not reproduce $f(E)$.

In particular, note that in eq.~\ref{eq:rho_and_fEL}, $\rho (r)$ is determined by an integral over
the range of $E$ and $L$ which are kinematically accessible at radial position $r$.  If $f(E,L)$ is largely independent
of $L$ at small $E$, then eq.~\ref{eq:rho_and_fE} will be a good approximation to $f(E)$ at small $E$.  On the other hand, the inverse Abel
integral transform in eq.~\ref{eq:Eddington} determines $f(E)$ in terms of an
integral over values of $\Phi$ (equivalently, $r$) which are not accessible
for a particle with energy $E$.  If $f$ is not a function of $E$ alone everywhere, this inverse equation need not exactly reproduce
$f(E)$, even at energies for which the velocity distribution is independent of $L$.  Further investigation of the slight discrepancy
between $f(E)$ obtained from numerical simulation data and from Eddington inversion, especially with a larger data set, would be
an interesting topic for future work.

\section{Conclusion}
\label{sec:conclusion}

We have compared the dark matter halo velocity distribution found in  DM-only numerical simulations
to analytic predictions, using the publicly-available Via Lactea 2 dataset as an example.
We have found them to be broadly consistent in the region lying well inside the scale radius.
In particular, we have found that the velocity
distribution found in numerical simulation is well described as a function of a single integral of motion --
the energy.  This is
consistent with analytic predictions in the case in which the dark matter distribution is spherically symmetric,
isotropic and time-invariant.  More specifically, the velocity distribution obtained from numerical
simulation data is a good
fit to the result obtained from Eddington inversion, well inside the scale radius.

Of course, there are certainly deviations from spherical symmetry, isotropy and time-invariance~\cite{Zemp:2008gw,Vera-Ciro:2011asd},
and the velocity distribution
in a localized region will be significantly affected by the recent merger history~\cite{Diemand:2007qr,Vogelsberger:2008qb,Necib:2018iwb}, etc.   Such deviations are especially
important in the context of direct detection experiments, for which experimental sensitivity depends on the
dark matter velocity distribution at a particular location in the Milky Way halo.
But our result implies
that deviations from these approximations do not dramatically alter the broad features of the velocity distribution
which one obtains from analytic methods, when averaged over sufficiently large scales.  These features are more
important in the context of indirect detection, where one is interested in dark matter annihilation within an entire halo or
subhalo.

Deviations from spherical symmetry and time-invariance can certainly be important in the context of indirect
detection as well, particularly in relation to the formation and disruption of substructure~\cite{Ghigna:1998vn,Munoz:2007jn,Wang:2016qol}.
We have not attempted
to consider this issue here, though it would be an interesting topic of future work.
However, we have considered the
application of our results to the case of dark matter annihilation from a $p$- or $d$-wave initial state, for which the
form of the velocity distribution is important.  In these scenarios, the effect of substructure is expected to be relatively mild~\cite{Baxter:2022dpn}.

Because  the $p$-wave annihilation cross section scales as $(v/c)^2$,
the contribution of the high-speed tail of the velocity
distribution is enhanced, and it becomes important to know how strongly the velocity distribution is
suppressed at high speed.
Since the tail of the distribution will be least well-sampled in a numerical simulation, it is helpful to
be able to gain intuition from analytic results.
If the velocity distribution is a function of energy alone, then within a power-law cusp,
the velocity distribution will generally fall off only as a power of velocity~\cite{Boucher:2021mii}, not exponentially (as one would
expect if the velocity distribution were Maxwell-Boltzmann).
To characterize the effect on $p$-wave annihilation of using a velocity distribution derived from
analytic principles, as opposed to a Maxwell-Boltzmann distribution, it is sufficient to compute the
velocity dispersion, $\langle v^2 \rangle$ for each scenario, since the annihilation rate per volume in the
case of $p$-wave annihilation is proportional to $\langle v^2 \rangle$~\cite{Boucher:2021mii}.  We compare
the result obtained from Eddington inversion deep inside a cusp with $\gamma = 1.24$ (where $f$ can be well approximated as
power law in $E$)  to a Maxwell-Boltzman distribution normalized so that both distributions
have the same peak velocity.  We find $\langle v^2 \rangle_{Edd.} / \langle v^2 \rangle_{MB}= 4.16$, indicating
that the exponential suppression of the high-speed tail in the Maxwell-Boltzmann distribution has a significant
effect on the $p$-wave annihilation rate,  As we find that numerical simulation data is well described by
the Eddington inversion result, at least for DM-only simulations, well inside the cusp, our results suggest that
$p$-wave annihilation may be significantly enhanced relative to expectations from a Maxwell-Boltzmann distribution.

Indeed, these results suggest that for a dark matter
halo of gNFW form with $\gamma = 1.24$ (as in the VL-2 halo), $d$-wave annihilation (which scales as
$(v/c)^4$) even within the cusp is dominated by the most energetic particles, which explore the entirety of the
halo~\cite{Boucher:2021mii}.  In this case, the total annihilation rate is controlled by the shape of the gravitational potential
well outside the scale radius.  These results may also impact studies of direct detection for models in which
only the high-velocity tail can provide recoils which are above threshold.

Our analysis has been performed only with the $10^5$ VL-2 particle sample made publicly available.  It would
be interesting to refine this analysis by applying it to a much larger data set.  In such an analysis, the effects
of anisotropy may become noticeable, requiring one to consider a velocity distribution depending on angular momentum
as well as energy.  Moreover, it would be worthwhile to see if the effects of deviations from spherical symmetry and
isotropy  become more noticeable at larger distances.
VL-2 is a DM-only simulation of a Milky Way-sized halo.
It would be interesting to extend this analysis to simulations which include baryonic matter, and on different
scales.

{\bf Acknowledgements}  For facilitating portions of this research, JK and LES wish to acknowledge the Center for Theoretical Underground Physics and Related Areas (CETUP*), the Institute for Underground Science at Sanford Underground Research Facility (SURF), and the South Dakota Science and Technology Authority for hospitality and financial support, as well as for providing a stimulating environment.
JK is supported in part by DOE grant DE-SC0010504.
LES is supported in part by DOE grant DE-SC0010813.

\bibliography{thebib.bib}

\begin{thebibliography}{36}%
\makeatletter
\providecommand \@ifxundefined [1]{%
 \@ifx{#1\undefined}
}%
\providecommand \@ifnum [1]{%
 \ifnum #1\expandafter \@firstoftwo
 \else \expandafter \@secondoftwo
 \fi
}%
\providecommand \@ifx [1]{%
 \ifx #1\expandafter \@firstoftwo
 \else \expandafter \@secondoftwo
 \fi
}%
\providecommand \natexlab [1]{#1}%
\providecommand \enquote  [1]{``#1''}%
\providecommand \bibnamefont  [1]{#1}%
\providecommand \bibfnamefont [1]{#1}%
\providecommand \citenamefont [1]{#1}%
\providecommand \href@noop [0]{\@secondoftwo}%
\providecommand \href [0]{\begingroup \@sanitize@url \@href}%
\providecommand \@href[1]{\@@startlink{#1}\@@href}%
\providecommand \@@href[1]{\endgroup#1\@@endlink}%
\providecommand \@sanitize@url [0]{\catcode `\\12\catcode `\$12\catcode
  `\&12\catcode `\#12\catcode `\^12\catcode `\_12\catcode `\%12\relax}%
\providecommand \@@startlink[1]{}%
\providecommand \@@endlink[0]{}%
\providecommand \url  [0]{\begingroup\@sanitize@url \@url }%
\providecommand \@url [1]{\endgroup\@href {#1}{\urlprefix }}%
\providecommand \urlprefix  [0]{URL }%
\providecommand \Eprint [0]{\href }%
\providecommand \doibase [0]{https://doi.org/}%
\providecommand \selectlanguage [0]{\@gobble}%
\providecommand \bibinfo  [0]{\@secondoftwo}%
\providecommand \bibfield  [0]{\@secondoftwo}%
\providecommand \translation [1]{[#1]}%
\providecommand \BibitemOpen [0]{}%
\providecommand \bibitemStop [0]{}%
\providecommand \bibitemNoStop [0]{.\EOS\space}%
\providecommand \EOS [0]{\spacefactor3000\relax}%
\providecommand \BibitemShut  [1]{\csname bibitem#1\endcsname}%
\let\auto@bib@innerbib\@empty
\bibitem [{\citenamefont {Diemand}\ \emph
  {et~al.}(2007{\natexlab{a}})\citenamefont {Diemand}, \citenamefont {Kuhlen},\
  and\ \citenamefont {Madau}}]{Diemand:2006ik}%
  \BibitemOpen
  \bibfield  {author} {\bibinfo {author} {\bibfnamefont {J.}~\bibnamefont
  {Diemand}}, \bibinfo {author} {\bibfnamefont {M.}~\bibnamefont {Kuhlen}},\
  and\ \bibinfo {author} {\bibfnamefont {P.}~\bibnamefont {Madau}},\ }\bibfield
   {title} {\bibinfo {title} {{Dark matter substructure and gamma-ray
  annihilation in the Milky Way halo}},\ }\href
  {https://doi.org/10.1086/510736} {\bibfield  {journal} {\bibinfo  {journal}
  {Astrophys. J.}\ }\textbf {\bibinfo {volume} {657}},\ \bibinfo {pages} {262}
  (\bibinfo {year} {2007}{\natexlab{a}})},\ \Eprint
  {https://arxiv.org/abs/astro-ph/0611370} {arXiv:astro-ph/0611370}
  \BibitemShut {NoStop}%
\bibitem [{\citenamefont {Vogelsberger}\ \emph {et~al.}(2008)\citenamefont
  {Vogelsberger}, \citenamefont {White}, \citenamefont {Helmi},\ and\
  \citenamefont {Springel}}]{Vogelsberger:2007ny}%
  \BibitemOpen
  \bibfield  {author} {\bibinfo {author} {\bibfnamefont {M.}~\bibnamefont
  {Vogelsberger}}, \bibinfo {author} {\bibfnamefont {S.~D.~M.}\ \bibnamefont
  {White}}, \bibinfo {author} {\bibfnamefont {A.}~\bibnamefont {Helmi}},\ and\
  \bibinfo {author} {\bibfnamefont {V.}~\bibnamefont {Springel}},\ }\bibfield
  {title} {\bibinfo {title} {{The fine-grained phase-space structure of Cold
  Dark Matter halos}},\ }\href
  {https://doi.org/10.1111/j.1365-2966.2007.12746.x} {\bibfield  {journal}
  {\bibinfo  {journal} {Mon. Not. Roy. Astron. Soc.}\ }\textbf {\bibinfo
  {volume} {385}},\ \bibinfo {pages} {236} (\bibinfo {year} {2008})},\ \Eprint
  {https://arxiv.org/abs/0711.1105} {arXiv:0711.1105 [astro-ph]} \BibitemShut
  {NoStop}%
\bibitem [{\citenamefont {Diemand}\ \emph {et~al.}(2008)\citenamefont
  {Diemand}, \citenamefont {Kuhlen}, \citenamefont {Madau}, \citenamefont
  {Zemp}, \citenamefont {Moore}, \citenamefont {Potter},\ and\ \citenamefont
  {Stadel}}]{Diemand:2008in}%
  \BibitemOpen
  \bibfield  {author} {\bibinfo {author} {\bibfnamefont {J.}~\bibnamefont
  {Diemand}}, \bibinfo {author} {\bibfnamefont {M.}~\bibnamefont {Kuhlen}},
  \bibinfo {author} {\bibfnamefont {P.}~\bibnamefont {Madau}}, \bibinfo
  {author} {\bibfnamefont {M.}~\bibnamefont {Zemp}}, \bibinfo {author}
  {\bibfnamefont {B.}~\bibnamefont {Moore}}, \bibinfo {author} {\bibfnamefont
  {D.}~\bibnamefont {Potter}},\ and\ \bibinfo {author} {\bibfnamefont
  {J.}~\bibnamefont {Stadel}},\ }\bibfield  {title} {\bibinfo {title} {{Clumps
  and streams in the local dark matter distribution}},\ }\href
  {https://doi.org/10.1038/nature07153} {\bibfield  {journal} {\bibinfo
  {journal} {Nature}\ }\textbf {\bibinfo {volume} {454}},\ \bibinfo {pages}
  {735} (\bibinfo {year} {2008})},\ \Eprint {https://arxiv.org/abs/0805.1244}
  {arXiv:0805.1244 [astro-ph]} \BibitemShut {NoStop}%
\bibitem [{\citenamefont {Stadel}\ \emph {et~al.}(2009)\citenamefont {Stadel},
  \citenamefont {Potter}, \citenamefont {Moore}, \citenamefont {Diemand},
  \citenamefont {Madau}, \citenamefont {Zemp}, \citenamefont {Kuhlen},\ and\
  \citenamefont {Quilis}}]{Stadel:2008pn}%
  \BibitemOpen
  \bibfield  {author} {\bibinfo {author} {\bibfnamefont {J.}~\bibnamefont
  {Stadel}}, \bibinfo {author} {\bibfnamefont {D.}~\bibnamefont {Potter}},
  \bibinfo {author} {\bibfnamefont {B.}~\bibnamefont {Moore}}, \bibinfo
  {author} {\bibfnamefont {J.}~\bibnamefont {Diemand}}, \bibinfo {author}
  {\bibfnamefont {P.}~\bibnamefont {Madau}}, \bibinfo {author} {\bibfnamefont
  {M.}~\bibnamefont {Zemp}}, \bibinfo {author} {\bibfnamefont {M.}~\bibnamefont
  {Kuhlen}},\ and\ \bibinfo {author} {\bibfnamefont {V.}~\bibnamefont
  {Quilis}},\ }\bibfield  {title} {\bibinfo {title} {{Quantifying the heart of
  darkness with GHALO - a multi-billion particle simulation of our galactic
  halo}},\ }\href {https://doi.org/10.1111/j.1745-3933.2009.00699.x} {\bibfield
   {journal} {\bibinfo  {journal} {Mon. Not. Roy. Astron. Soc.}\ }\textbf
  {\bibinfo {volume} {398}},\ \bibinfo {pages} {L21} (\bibinfo {year}
  {2009})},\ \Eprint {https://arxiv.org/abs/0808.2981} {arXiv:0808.2981
  [astro-ph]} \BibitemShut {NoStop}%
\bibitem [{\citenamefont {Widrow}(2000)}]{Widrow:2000dm}%
  \BibitemOpen
  \bibfield  {author} {\bibinfo {author} {\bibfnamefont {L.~M.}\ \bibnamefont
  {Widrow}},\ }\bibfield  {title} {\bibinfo {title} {{Semi-analytic models for
  dark matter halos}},\ }\href@noop {} {\  (\bibinfo {year} {2000})},\ \Eprint
  {https://arxiv.org/abs/astro-ph/0003302} {arXiv:astro-ph/0003302}
  \BibitemShut {NoStop}%
\bibitem [{\citenamefont {Evans}\ and\ \citenamefont
  {An}(2006)}]{Evans:2005tn}%
  \BibitemOpen
  \bibfield  {author} {\bibinfo {author} {\bibfnamefont {N.~W.}\ \bibnamefont
  {Evans}}\ and\ \bibinfo {author} {\bibfnamefont {J.~H.}\ \bibnamefont {An}},\
  }\bibfield  {title} {\bibinfo {title} {{Distribution function of the dark
  matter}},\ }\href {https://doi.org/10.1103/PhysRevD.73.023524} {\bibfield
  {journal} {\bibinfo  {journal} {Phys. Rev. D}\ }\textbf {\bibinfo {volume}
  {73}},\ \bibinfo {pages} {023524} (\bibinfo {year} {2006})},\ \Eprint
  {https://arxiv.org/abs/astro-ph/0511687} {arXiv:astro-ph/0511687}
  \BibitemShut {NoStop}%
\bibitem [{\citenamefont {Wojtak}\ \emph {et~al.}(2008)\citenamefont {Wojtak},
  \citenamefont {Lokas}, \citenamefont {Mamon}, \citenamefont {Gottloeber},
  \citenamefont {Klypin},\ and\ \citenamefont {Hoffman}}]{Wojtak:2008mg}%
  \BibitemOpen
  \bibfield  {author} {\bibinfo {author} {\bibfnamefont {R.}~\bibnamefont
  {Wojtak}}, \bibinfo {author} {\bibfnamefont {E.~L.}\ \bibnamefont {Lokas}},
  \bibinfo {author} {\bibfnamefont {G.~A.}\ \bibnamefont {Mamon}}, \bibinfo
  {author} {\bibfnamefont {S.}~\bibnamefont {Gottloeber}}, \bibinfo {author}
  {\bibfnamefont {A.}~\bibnamefont {Klypin}},\ and\ \bibinfo {author}
  {\bibfnamefont {Y.}~\bibnamefont {Hoffman}},\ }\bibfield  {title} {\bibinfo
  {title} {{The distribution function of dark matter in massive haloes}},\
  }\href {https://doi.org/10.1111/j.1365-2966.2008.13441.x} {\bibfield
  {journal} {\bibinfo  {journal} {Mon. Not. Roy. Astron. Soc.}\ }\textbf
  {\bibinfo {volume} {388}},\ \bibinfo {pages} {815} (\bibinfo {year}
  {2008})},\ \Eprint {https://arxiv.org/abs/0802.0429} {arXiv:0802.0429
  [astro-ph]} \BibitemShut {NoStop}%
\bibitem [{\citenamefont {Mamon}\ \emph {et~al.}(2013)\citenamefont {Mamon},
  \citenamefont {Biviano},\ and\ \citenamefont {Boue}}]{Mamon:2012yb}%
  \BibitemOpen
  \bibfield  {author} {\bibinfo {author} {\bibfnamefont {G.~A.}\ \bibnamefont
  {Mamon}}, \bibinfo {author} {\bibfnamefont {A.}~\bibnamefont {Biviano}},\
  and\ \bibinfo {author} {\bibfnamefont {G.}~\bibnamefont {Boue}},\ }\bibfield
  {title} {\bibinfo {title} {{MAMPOSSt: Modelling Anisotropy and Mass Profiles
  of Observed Spherical Systems. I. Gaussian 3D velocities}},\ }\href
  {https://doi.org/10.1093/mnras/sts565} {\bibfield  {journal} {\bibinfo
  {journal} {Mon. Not. Roy. Astron. Soc.}\ }\textbf {\bibinfo {volume} {429}},\
  \bibinfo {pages} {3079} (\bibinfo {year} {2013})},\ \Eprint
  {https://arxiv.org/abs/1212.1455} {arXiv:1212.1455 [astro-ph.CO]}
  \BibitemShut {NoStop}%
\bibitem [{\citenamefont {Peta\v{c}}\ \emph {et~al.}(2021)\citenamefont
  {Peta\v{c}}, \citenamefont {Lavalle}, \citenamefont {N\'u\~nez
  Casti\~neyra},\ and\ \citenamefont {Nezri}}]{Petac:2021cjm}%
  \BibitemOpen
  \bibfield  {author} {\bibinfo {author} {\bibfnamefont {M.}~\bibnamefont
  {Peta\v{c}}}, \bibinfo {author} {\bibfnamefont {J.}~\bibnamefont {Lavalle}},
  \bibinfo {author} {\bibfnamefont {A.}~\bibnamefont {N\'u\~nez
  Casti\~neyra}},\ and\ \bibinfo {author} {\bibfnamefont {E.}~\bibnamefont
  {Nezri}},\ }\bibfield  {title} {\bibinfo {title} {{Testing the predictions of
  axisymmetric distribution functions of galactic dark matter with
  hydrodynamical simulations}},\ }\href
  {https://doi.org/10.1088/1475-7516/2021/08/031} {\bibfield  {journal}
  {\bibinfo  {journal} {JCAP}\ }\textbf {\bibinfo {volume} {08}},\ \bibinfo
  {pages} {031}},\ \Eprint {https://arxiv.org/abs/2106.01314} {arXiv:2106.01314
  [astro-ph.CO]} \BibitemShut {NoStop}%
\bibitem [{\citenamefont {Fox}\ \emph {et~al.}(2011)\citenamefont {Fox},
  \citenamefont {Kribs},\ and\ \citenamefont {Tait}}]{Fox:2010bu}%
  \BibitemOpen
  \bibfield  {author} {\bibinfo {author} {\bibfnamefont {P.~J.}\ \bibnamefont
  {Fox}}, \bibinfo {author} {\bibfnamefont {G.~D.}\ \bibnamefont {Kribs}},\
  and\ \bibinfo {author} {\bibfnamefont {T.~M.~P.}\ \bibnamefont {Tait}},\
  }\bibfield  {title} {\bibinfo {title} {{Interpreting Dark Matter Direct
  Detection Independently of the Local Velocity and Density Distribution}},\
  }\href {https://doi.org/10.1103/PhysRevD.83.034007} {\bibfield  {journal}
  {\bibinfo  {journal} {Phys. Rev. D}\ }\textbf {\bibinfo {volume} {83}},\
  \bibinfo {pages} {034007} (\bibinfo {year} {2011})},\ \Eprint
  {https://arxiv.org/abs/1011.1910} {arXiv:1011.1910 [hep-ph]} \BibitemShut
  {NoStop}%
\bibitem [{\citenamefont {Freese}\ \emph {et~al.}(2013)\citenamefont {Freese},
  \citenamefont {Lisanti},\ and\ \citenamefont {Savage}}]{Freese:2012xd}%
  \BibitemOpen
  \bibfield  {author} {\bibinfo {author} {\bibfnamefont {K.}~\bibnamefont
  {Freese}}, \bibinfo {author} {\bibfnamefont {M.}~\bibnamefont {Lisanti}},\
  and\ \bibinfo {author} {\bibfnamefont {C.}~\bibnamefont {Savage}},\
  }\bibfield  {title} {\bibinfo {title} {{Colloquium: Annual modulation of dark
  matter}},\ }\href {https://doi.org/10.1103/RevModPhys.85.1561} {\bibfield
  {journal} {\bibinfo  {journal} {Rev. Mod. Phys.}\ }\textbf {\bibinfo {volume}
  {85}},\ \bibinfo {pages} {1561} (\bibinfo {year} {2013})},\ \Eprint
  {https://arxiv.org/abs/1209.3339} {arXiv:1209.3339 [astro-ph.CO]}
  \BibitemShut {NoStop}%
\bibitem [{\citenamefont {{Robertson}}\ and\ \citenamefont
  {{Zentner}}(2009)}]{2009PhRvD..79h3525R}%
  \BibitemOpen
  \bibfield  {author} {\bibinfo {author} {\bibfnamefont {B.~E.}\ \bibnamefont
  {{Robertson}}}\ and\ \bibinfo {author} {\bibfnamefont {A.~R.}\ \bibnamefont
  {{Zentner}}},\ }\bibfield  {title} {\bibinfo {title} {{Dark matter
  annihilation rates with velocity-dependent annihilation cross sections}},\
  }\href {https://doi.org/10.1103/PhysRevD.79.083525} {\bibfield  {journal}
  {\bibinfo  {journal} {\prd}\ }\textbf {\bibinfo {volume} {79}},\ \bibinfo
  {eid} {083525} (\bibinfo {year} {2009})},\ \Eprint
  {https://arxiv.org/abs/0902.0362} {arXiv:0902.0362 [astro-ph.CO]}
  \BibitemShut {NoStop}%
\bibitem [{\citenamefont {{Belotsky}}\ \emph {et~al.}(2014)\citenamefont
  {{Belotsky}}, \citenamefont {{Kirillov}},\ and\ \citenamefont
  {{Khlopov}}}]{2014GrCo...20...47B}%
  \BibitemOpen
  \bibfield  {author} {\bibinfo {author} {\bibfnamefont {K.}~\bibnamefont
  {{Belotsky}}}, \bibinfo {author} {\bibfnamefont {A.}~\bibnamefont
  {{Kirillov}}},\ and\ \bibinfo {author} {\bibfnamefont {M.}~\bibnamefont
  {{Khlopov}}},\ }\bibfield  {title} {\bibinfo {title} {{Gamma-ray evidence for
  dark matter clumps}},\ }\href {https://doi.org/10.1134/S0202289314010022}
  {\bibfield  {journal} {\bibinfo  {journal} {Gravitation and Cosmology}\
  }\textbf {\bibinfo {volume} {20}},\ \bibinfo {pages} {47} (\bibinfo {year}
  {2014})},\ \Eprint {https://arxiv.org/abs/1212.6087} {arXiv:1212.6087
  [astro-ph.HE]} \BibitemShut {NoStop}%
\bibitem [{\citenamefont {Ferrer}\ and\ \citenamefont
  {Hunter}(2013)}]{Ferrer:2013cla}%
  \BibitemOpen
  \bibfield  {author} {\bibinfo {author} {\bibfnamefont {F.}~\bibnamefont
  {Ferrer}}\ and\ \bibinfo {author} {\bibfnamefont {D.~R.}\ \bibnamefont
  {Hunter}},\ }\bibfield  {title} {\bibinfo {title} {{The impact of the
  phase-space density on the indirect detection of dark matter}},\ }\href
  {https://doi.org/10.1088/1475-7516/2013/09/005} {\bibfield  {journal}
  {\bibinfo  {journal} {JCAP}\ }\textbf {\bibinfo {volume} {09}},\ \bibinfo
  {pages} {005}},\ \Eprint {https://arxiv.org/abs/1306.6586} {arXiv:1306.6586
  [astro-ph.HE]} \BibitemShut {NoStop}%
\bibitem [{\citenamefont {Boddy}\ \emph {et~al.}(2017)\citenamefont {Boddy},
  \citenamefont {Kumar}, \citenamefont {Strigari},\ and\ \citenamefont
  {Wang}}]{Boddy:2017vpe}%
  \BibitemOpen
  \bibfield  {author} {\bibinfo {author} {\bibfnamefont {K.~K.}\ \bibnamefont
  {Boddy}}, \bibinfo {author} {\bibfnamefont {J.}~\bibnamefont {Kumar}},
  \bibinfo {author} {\bibfnamefont {L.~E.}\ \bibnamefont {Strigari}},\ and\
  \bibinfo {author} {\bibfnamefont {M.-Y.}\ \bibnamefont {Wang}},\ }\bibfield
  {title} {\bibinfo {title} {{Sommerfeld-Enhanced $J$-Factors For Dwarf
  Spheroidal Galaxies}},\ }\href {https://doi.org/10.1103/PhysRevD.95.123008}
  {\bibfield  {journal} {\bibinfo  {journal} {Phys. Rev. D}\ }\textbf {\bibinfo
  {volume} {95}},\ \bibinfo {pages} {123008} (\bibinfo {year} {2017})},\
  \Eprint {https://arxiv.org/abs/1702.00408} {arXiv:1702.00408 [astro-ph.CO]}
  \BibitemShut {NoStop}%
\bibitem [{\citenamefont {Piccirillo}\ \emph {et~al.}(2022)\citenamefont
  {Piccirillo}, \citenamefont {Blanchette}, \citenamefont {Bozorgnia},
  \citenamefont {Strigari}, \citenamefont {Frenk}, \citenamefont {Grand},\ and\
  \citenamefont {Marinacci}}]{Piccirillo:2022qet}%
  \BibitemOpen
  \bibfield  {author} {\bibinfo {author} {\bibfnamefont {E.}~\bibnamefont
  {Piccirillo}}, \bibinfo {author} {\bibfnamefont {K.}~\bibnamefont
  {Blanchette}}, \bibinfo {author} {\bibfnamefont {N.}~\bibnamefont
  {Bozorgnia}}, \bibinfo {author} {\bibfnamefont {L.~E.}\ \bibnamefont
  {Strigari}}, \bibinfo {author} {\bibfnamefont {C.~S.}\ \bibnamefont {Frenk}},
  \bibinfo {author} {\bibfnamefont {R.~J.~J.}\ \bibnamefont {Grand}},\ and\
  \bibinfo {author} {\bibfnamefont {F.}~\bibnamefont {Marinacci}},\ }\bibfield
  {title} {\bibinfo {title} {{Velocity-dependent annihilation radiation from
  dark matter subhalos in cosmological simulations}},\ }\href
  {https://doi.org/10.1088/1475-7516/2022/08/058} {\bibfield  {journal}
  {\bibinfo  {journal} {JCAP}\ }\textbf {\bibinfo {volume} {08}}\bibfield
  {number} {\bibinfo  {number} { (08)},\ \bibinfo {pages} {058}},\ }\Eprint
  {https://arxiv.org/abs/2203.08853} {arXiv:2203.08853 [astro-ph.CO]}
  \BibitemShut {NoStop}%
\bibitem [{\citenamefont {Blanchette}\ \emph {et~al.}(2023)\citenamefont
  {Blanchette}, \citenamefont {Piccirillo}, \citenamefont {Bozorgnia},
  \citenamefont {Strigari}, \citenamefont {Fattahi}, \citenamefont {Frenk},
  \citenamefont {Navarro},\ and\ \citenamefont {Sawala}}]{Blanchette:2022hir}%
  \BibitemOpen
  \bibfield  {author} {\bibinfo {author} {\bibfnamefont {K.}~\bibnamefont
  {Blanchette}}, \bibinfo {author} {\bibfnamefont {E.}~\bibnamefont
  {Piccirillo}}, \bibinfo {author} {\bibfnamefont {N.}~\bibnamefont
  {Bozorgnia}}, \bibinfo {author} {\bibfnamefont {L.~E.}\ \bibnamefont
  {Strigari}}, \bibinfo {author} {\bibfnamefont {A.}~\bibnamefont {Fattahi}},
  \bibinfo {author} {\bibfnamefont {C.~S.}\ \bibnamefont {Frenk}}, \bibinfo
  {author} {\bibfnamefont {J.~F.}\ \bibnamefont {Navarro}},\ and\ \bibinfo
  {author} {\bibfnamefont {T.}~\bibnamefont {Sawala}},\ }\bibfield  {title}
  {\bibinfo {title} {{Velocity-dependent J-factors for Milky Way dwarf
  spheroidal analogues in cosmological simulations}},\ }\href
  {https://doi.org/10.1088/1475-7516/2023/03/021} {\bibfield  {journal}
  {\bibinfo  {journal} {JCAP}\ }\textbf {\bibinfo {volume} {03}},\ \bibinfo
  {pages} {021}},\ \Eprint {https://arxiv.org/abs/2207.00069} {arXiv:2207.00069
  [astro-ph.CO]} \BibitemShut {NoStop}%
\bibitem [{\citenamefont {Kuhlen}\ \emph {et~al.}(2010)\citenamefont {Kuhlen},
  \citenamefont {Weiner}, \citenamefont {Diemand}, \citenamefont {Madau},
  \citenamefont {Moore}, \citenamefont {Potter}, \citenamefont {Stadel},\ and\
  \citenamefont {Zemp}}]{Kuhlen:2009vh}%
  \BibitemOpen
  \bibfield  {author} {\bibinfo {author} {\bibfnamefont {M.}~\bibnamefont
  {Kuhlen}}, \bibinfo {author} {\bibfnamefont {N.}~\bibnamefont {Weiner}},
  \bibinfo {author} {\bibfnamefont {J.}~\bibnamefont {Diemand}}, \bibinfo
  {author} {\bibfnamefont {P.}~\bibnamefont {Madau}}, \bibinfo {author}
  {\bibfnamefont {B.}~\bibnamefont {Moore}}, \bibinfo {author} {\bibfnamefont
  {D.}~\bibnamefont {Potter}}, \bibinfo {author} {\bibfnamefont
  {J.}~\bibnamefont {Stadel}},\ and\ \bibinfo {author} {\bibfnamefont
  {M.}~\bibnamefont {Zemp}},\ }\bibfield  {title} {\bibinfo {title} {{Dark
  Matter Direct Detection with Non-Maxwellian Velocity Structure}},\ }\href
  {https://doi.org/10.1088/1475-7516/2010/02/030} {\bibfield  {journal}
  {\bibinfo  {journal} {JCAP}\ }\textbf {\bibinfo {volume} {02}},\ \bibinfo
  {pages} {030}},\ \Eprint {https://arxiv.org/abs/0912.2358} {arXiv:0912.2358
  [astro-ph.GA]} \BibitemShut {NoStop}%
\bibitem [{\citenamefont {Beraldo~e Silva}\ \emph {et~al.}(2015)\citenamefont
  {Beraldo~e Silva}, \citenamefont {Mamon}, \citenamefont {Duarte},
  \citenamefont {Wojtak}, \citenamefont {Peirani},\ and\ \citenamefont
  {Bou\'e}}]{BeraldoeSilva:2013efk}%
  \BibitemOpen
  \bibfield  {author} {\bibinfo {author} {\bibfnamefont {L.}~\bibnamefont
  {Beraldo~e Silva}}, \bibinfo {author} {\bibfnamefont {G.~A.}\ \bibnamefont
  {Mamon}}, \bibinfo {author} {\bibfnamefont {M.}~\bibnamefont {Duarte}},
  \bibinfo {author} {\bibfnamefont {R.}~\bibnamefont {Wojtak}}, \bibinfo
  {author} {\bibfnamefont {S.}~\bibnamefont {Peirani}},\ and\ \bibinfo {author}
  {\bibfnamefont {G.}~\bibnamefont {Bou\'e}},\ }\bibfield  {title} {\bibinfo
  {title} {{Anisotropic $q$-Gaussian 3D velocity distributions in
  \ensuremath{\Lambda}CDM haloes}},\ }\href
  {https://doi.org/10.1093/mnras/stv1321} {\bibfield  {journal} {\bibinfo
  {journal} {Mon. Not. Roy. Astron. Soc.}\ }\textbf {\bibinfo {volume} {452}},\
  \bibinfo {pages} {944} (\bibinfo {year} {2015})},\ \bibinfo {note} {[Erratum:
  Mon.Not.Roy.Astron.Soc. 467, 2445 (2017)]},\ \Eprint
  {https://arxiv.org/abs/1310.6756} {arXiv:1310.6756 [astro-ph.CO]}
  \BibitemShut {NoStop}%
\bibitem [{\citenamefont {Boucher}\ \emph {et~al.}(2022)\citenamefont
  {Boucher}, \citenamefont {Kumar}, \citenamefont {Le},\ and\ \citenamefont
  {Runburg}}]{Boucher:2021mii}%
  \BibitemOpen
  \bibfield  {author} {\bibinfo {author} {\bibfnamefont {B.}~\bibnamefont
  {Boucher}}, \bibinfo {author} {\bibfnamefont {J.}~\bibnamefont {Kumar}},
  \bibinfo {author} {\bibfnamefont {V.~B.}\ \bibnamefont {Le}},\ and\ \bibinfo
  {author} {\bibfnamefont {J.}~\bibnamefont {Runburg}},\ }\bibfield  {title}
  {\bibinfo {title} {{J-factors for velocity-dependent dark matter
  annihilation}},\ }\href {https://doi.org/10.1103/PhysRevD.106.023025}
  {\bibfield  {journal} {\bibinfo  {journal} {Phys. Rev. D}\ }\textbf {\bibinfo
  {volume} {106}},\ \bibinfo {pages} {023025} (\bibinfo {year} {2022})},\
  \Eprint {https://arxiv.org/abs/2110.09653} {arXiv:2110.09653 [hep-ph]}
  \BibitemShut {NoStop}%
\bibitem [{\citenamefont {Kiriu}\ \emph {et~al.}(2022)\citenamefont {Kiriu},
  \citenamefont {Kumar},\ and\ \citenamefont {Runburg}}]{Kiriu:2022bjq}%
  \BibitemOpen
  \bibfield  {author} {\bibinfo {author} {\bibfnamefont {K.}~\bibnamefont
  {Kiriu}}, \bibinfo {author} {\bibfnamefont {J.}~\bibnamefont {Kumar}},\ and\
  \bibinfo {author} {\bibfnamefont {J.}~\bibnamefont {Runburg}},\ }\bibfield
  {title} {\bibinfo {title} {{The velocity-dependent J-factor of the Milky Way
  halo: does what happens in the galactic bulge stay in the galactic bulge?}},\
  }\href {https://doi.org/10.1088/1475-7516/2022/11/030} {\bibfield  {journal}
  {\bibinfo  {journal} {JCAP}\ }\textbf {\bibinfo {volume} {11}},\ \bibinfo
  {pages} {030}},\ \Eprint {https://arxiv.org/abs/2208.14002} {arXiv:2208.14002
  [hep-ph]} \BibitemShut {NoStop}%
\bibitem [{\citenamefont {Lacroix}\ \emph {et~al.}(2020)\citenamefont
  {Lacroix}, \citenamefont {N\'u\~nez Casti\~neyra}, \citenamefont {Stref},
  \citenamefont {Lavalle},\ and\ \citenamefont {Nezri}}]{Lacroix:2020lhn}%
  \BibitemOpen
  \bibfield  {author} {\bibinfo {author} {\bibfnamefont {T.}~\bibnamefont
  {Lacroix}}, \bibinfo {author} {\bibfnamefont {A.}~\bibnamefont {N\'u\~nez
  Casti\~neyra}}, \bibinfo {author} {\bibfnamefont {M.}~\bibnamefont {Stref}},
  \bibinfo {author} {\bibfnamefont {J.}~\bibnamefont {Lavalle}},\ and\ \bibinfo
  {author} {\bibfnamefont {E.}~\bibnamefont {Nezri}},\ }\bibfield  {title}
  {\bibinfo {title} {{Predicting the dark matter velocity distribution in
  galactic structures: tests against hydrodynamic cosmological simulations}},\
  }\href {https://doi.org/10.1088/1475-7516/2020/10/031} {\bibfield  {journal}
  {\bibinfo  {journal} {JCAP}\ }\textbf {\bibinfo {volume} {10}},\ \bibinfo
  {pages} {031}},\ \Eprint {https://arxiv.org/abs/2005.03955} {arXiv:2005.03955
  [astro-ph.GA]} \BibitemShut {NoStop}%
\bibitem [{\citenamefont {Goldstein}(1980)}]{goldstein:mechanics}%
  \BibitemOpen
  \bibfield  {author} {\bibinfo {author} {\bibfnamefont {H.}~\bibnamefont
  {Goldstein}},\ }\href@noop {} {\emph {\bibinfo {title} {Classical
  Mechanics}}}\ (\bibinfo  {publisher} {Addison-Wesley},\ \bibinfo {year}
  {1980})\BibitemShut {NoStop}%
\bibitem [{\citenamefont {{Binney}}\ and\ \citenamefont
  {{Tremaine}}(2008)}]{2008gady.bookB}%
  \BibitemOpen
  \bibfield  {author} {\bibinfo {author} {\bibfnamefont {J.}~\bibnamefont
  {{Binney}}}\ and\ \bibinfo {author} {\bibfnamefont {S.}~\bibnamefont
  {{Tremaine}}},\ }\href@noop {} {\emph {\bibinfo {title} {{Galactic Dynamics:
  Second Edition}}}}\ (\bibinfo {year} {2008})\BibitemShut {NoStop}%
\bibitem [{\citenamefont {Zemp}(2009)}]{Zemp:2009ff}%
  \BibitemOpen
  \bibfield  {author} {\bibinfo {author} {\bibfnamefont {M.}~\bibnamefont
  {Zemp}},\ }\bibfield  {title} {\bibinfo {title} {{The Structure of Cold Dark
  Matter Halos: Recent Insights from High Resolution Simulations}},\ }\href
  {https://doi.org/10.1142/S0217732309031880} {\bibfield  {journal} {\bibinfo
  {journal} {Mod. Phys. Lett. A}\ }\textbf {\bibinfo {volume} {24}},\ \bibinfo
  {pages} {2291} (\bibinfo {year} {2009})},\ \Eprint
  {https://arxiv.org/abs/0909.4298} {arXiv:0909.4298 [astro-ph.CO]}
  \BibitemShut {NoStop}%
\bibitem [{\citenamefont {Zhao}(1997)}]{Zhao:1996mr}%
  \BibitemOpen
  \bibfield  {author} {\bibinfo {author} {\bibfnamefont {H.}~\bibnamefont
  {Zhao}},\ }\bibfield  {title} {\bibinfo {title} {{Analytical dynamical models
  for double-power-law galactic nuclei}},\ }\href
  {https://doi.org/10.1093/mnras/287.3.525} {\bibfield  {journal} {\bibinfo
  {journal} {Mon. Not. Roy. Astron. Soc.}\ }\textbf {\bibinfo {volume} {287}},\
  \bibinfo {pages} {525} (\bibinfo {year} {1997})},\ \Eprint
  {https://arxiv.org/abs/astro-ph/9605029} {arXiv:astro-ph/9605029}
  \BibitemShut {NoStop}%
\bibitem [{\citenamefont {Wojtak}\ \emph {et~al.}(2013)\citenamefont {Wojtak},
  \citenamefont {Gottloeber},\ and\ \citenamefont {Klypin}}]{Wojtak:2013eia}%
  \BibitemOpen
  \bibfield  {author} {\bibinfo {author} {\bibfnamefont {R.}~\bibnamefont
  {Wojtak}}, \bibinfo {author} {\bibfnamefont {S.}~\bibnamefont {Gottloeber}},\
  and\ \bibinfo {author} {\bibfnamefont {A.}~\bibnamefont {Klypin}},\
  }\bibfield  {title} {\bibinfo {title} {{Orbital anisotropy in cosmological
  haloes revisited}},\ }\href {https://doi.org/10.1093/mnras/stt1113}
  {\bibfield  {journal} {\bibinfo  {journal} {Mon. Not. Roy. Astron. Soc.}\
  }\textbf {\bibinfo {volume} {434}},\ \bibinfo {pages} {1576} (\bibinfo {year}
  {2013})},\ \Eprint {https://arxiv.org/abs/1303.2056} {arXiv:1303.2056
  [astro-ph.CO]} \BibitemShut {NoStop}%
\bibitem [{\citenamefont {Zemp}\ \emph {et~al.}(2009)\citenamefont {Zemp},
  \citenamefont {Diemand}, \citenamefont {Kuhlen}, \citenamefont {Madau},
  \citenamefont {Moore}, \citenamefont {Potter}, \citenamefont {Stadel},\ and\
  \citenamefont {Widrow}}]{Zemp:2008gw}%
  \BibitemOpen
  \bibfield  {author} {\bibinfo {author} {\bibfnamefont {M.}~\bibnamefont
  {Zemp}}, \bibinfo {author} {\bibfnamefont {J.}~\bibnamefont {Diemand}},
  \bibinfo {author} {\bibfnamefont {M.}~\bibnamefont {Kuhlen}}, \bibinfo
  {author} {\bibfnamefont {P.}~\bibnamefont {Madau}}, \bibinfo {author}
  {\bibfnamefont {B.}~\bibnamefont {Moore}}, \bibinfo {author} {\bibfnamefont
  {D.}~\bibnamefont {Potter}}, \bibinfo {author} {\bibfnamefont
  {J.}~\bibnamefont {Stadel}},\ and\ \bibinfo {author} {\bibfnamefont
  {L.}~\bibnamefont {Widrow}},\ }\bibfield  {title} {\bibinfo {title} {{The
  Graininess of Dark Matter Haloes}},\ }\href
  {https://doi.org/10.1111/j.1365-2966.2008.14361.x} {\bibfield  {journal}
  {\bibinfo  {journal} {Mon. Not. Roy. Astron. Soc.}\ }\textbf {\bibinfo
  {volume} {394}},\ \bibinfo {pages} {641} (\bibinfo {year} {2009})},\ \Eprint
  {https://arxiv.org/abs/0812.2033} {arXiv:0812.2033 [astro-ph]} \BibitemShut
  {NoStop}%
\bibitem [{\citenamefont {Vera-Ciro}\ \emph {et~al.}(2011)\citenamefont
  {Vera-Ciro}, \citenamefont {Sales}, \citenamefont {Helmi}, \citenamefont
  {Frenk}, \citenamefont {Navarro}, \citenamefont {Springel}, \citenamefont
  {Vogelsberger},\ and\ \citenamefont {White}}]{Vera-Ciro:2011asd}%
  \BibitemOpen
  \bibfield  {author} {\bibinfo {author} {\bibfnamefont {C.~A.}\ \bibnamefont
  {Vera-Ciro}}, \bibinfo {author} {\bibfnamefont {L.~V.}\ \bibnamefont
  {Sales}}, \bibinfo {author} {\bibfnamefont {A.}~\bibnamefont {Helmi}},
  \bibinfo {author} {\bibfnamefont {C.~S.}\ \bibnamefont {Frenk}}, \bibinfo
  {author} {\bibfnamefont {J.~F.}\ \bibnamefont {Navarro}}, \bibinfo {author}
  {\bibfnamefont {V.}~\bibnamefont {Springel}}, \bibinfo {author}
  {\bibfnamefont {M.}~\bibnamefont {Vogelsberger}},\ and\ \bibinfo {author}
  {\bibfnamefont {S.~D.~M.}\ \bibnamefont {White}},\ }\bibfield  {title}
  {\bibinfo {title} {{The Shape of Dark Matter Haloes in the Aquarius
  Simulations: Evolution and Memory}},\ }\href
  {https://doi.org/10.1111/j.1365-2966.2011.19134.x} {\bibfield  {journal}
  {\bibinfo  {journal} {Mon. Not. Roy. Astron. Soc.}\ }\textbf {\bibinfo
  {volume} {416}},\ \bibinfo {pages} {1377} (\bibinfo {year} {2011})},\ \Eprint
  {https://arxiv.org/abs/1104.1566} {arXiv:1104.1566 [astro-ph.CO]}
  \BibitemShut {NoStop}%
\bibitem [{\citenamefont {Diemand}\ \emph
  {et~al.}(2007{\natexlab{b}})\citenamefont {Diemand}, \citenamefont {Kuhlen},\
  and\ \citenamefont {Madau}}]{Diemand:2007qr}%
  \BibitemOpen
  \bibfield  {author} {\bibinfo {author} {\bibfnamefont {J.}~\bibnamefont
  {Diemand}}, \bibinfo {author} {\bibfnamefont {M.}~\bibnamefont {Kuhlen}},\
  and\ \bibinfo {author} {\bibfnamefont {P.}~\bibnamefont {Madau}},\ }\bibfield
   {title} {\bibinfo {title} {{Formation and evolution of galaxy dark matter
  halos and their substructure}},\ }\href {https://doi.org/10.1086/520573}
  {\bibfield  {journal} {\bibinfo  {journal} {Astrophys. J.}\ }\textbf
  {\bibinfo {volume} {667}},\ \bibinfo {pages} {859} (\bibinfo {year}
  {2007}{\natexlab{b}})},\ \Eprint {https://arxiv.org/abs/astro-ph/0703337}
  {arXiv:astro-ph/0703337} \BibitemShut {NoStop}%
\bibitem [{\citenamefont {Vogelsberger}\ \emph {et~al.}(2009)\citenamefont
  {Vogelsberger}, \citenamefont {Helmi}, \citenamefont {Springel},
  \citenamefont {White}, \citenamefont {Wang}, \citenamefont {Frenk},
  \citenamefont {Jenkins}, \citenamefont {Ludlow},\ and\ \citenamefont
  {Navarro}}]{Vogelsberger:2008qb}%
  \BibitemOpen
  \bibfield  {author} {\bibinfo {author} {\bibfnamefont {M.}~\bibnamefont
  {Vogelsberger}}, \bibinfo {author} {\bibfnamefont {A.}~\bibnamefont {Helmi}},
  \bibinfo {author} {\bibfnamefont {V.}~\bibnamefont {Springel}}, \bibinfo
  {author} {\bibfnamefont {S.~D.~M.}\ \bibnamefont {White}}, \bibinfo {author}
  {\bibfnamefont {J.}~\bibnamefont {Wang}}, \bibinfo {author} {\bibfnamefont
  {C.~S.}\ \bibnamefont {Frenk}}, \bibinfo {author} {\bibfnamefont
  {A.}~\bibnamefont {Jenkins}}, \bibinfo {author} {\bibfnamefont {A.~D.}\
  \bibnamefont {Ludlow}},\ and\ \bibinfo {author} {\bibfnamefont {J.~F.}\
  \bibnamefont {Navarro}},\ }\bibfield  {title} {\bibinfo {title} {{Phase-space
  structure in the local dark matter distribution and its signature in direct
  detection experiments}},\ }\href
  {https://doi.org/10.1111/j.1365-2966.2009.14630.x} {\bibfield  {journal}
  {\bibinfo  {journal} {Mon. Not. Roy. Astron. Soc.}\ }\textbf {\bibinfo
  {volume} {395}},\ \bibinfo {pages} {797} (\bibinfo {year} {2009})},\ \Eprint
  {https://arxiv.org/abs/0812.0362} {arXiv:0812.0362 [astro-ph]} \BibitemShut
  {NoStop}%
\bibitem [{\citenamefont {Necib}\ \emph {et~al.}(2018)\citenamefont {Necib},
  \citenamefont {Lisanti},\ and\ \citenamefont {Belokurov}}]{Necib:2018iwb}%
  \BibitemOpen
  \bibfield  {author} {\bibinfo {author} {\bibfnamefont {L.}~\bibnamefont
  {Necib}}, \bibinfo {author} {\bibfnamefont {M.}~\bibnamefont {Lisanti}},\
  and\ \bibinfo {author} {\bibfnamefont {V.}~\bibnamefont {Belokurov}},\
  }\bibfield  {title} {\bibinfo {title} {{Inferred Evidence For Dark Matter
  Kinematic Substructure with SDSS-Gaia}}\ }\href
  {https://doi.org/10.3847/1538-4357/ab095b} {10.3847/1538-4357/ab095b}
  (\bibinfo {year} {2018}),\ \Eprint {https://arxiv.org/abs/1807.02519}
  {arXiv:1807.02519 [astro-ph.GA]} \BibitemShut {NoStop}%
\bibitem [{\citenamefont {Ghigna}\ \emph {et~al.}(1998)\citenamefont {Ghigna},
  \citenamefont {Moore}, \citenamefont {Governato}, \citenamefont {Lake},
  \citenamefont {Quinn},\ and\ \citenamefont {Stadel}}]{Ghigna:1998vn}%
  \BibitemOpen
  \bibfield  {author} {\bibinfo {author} {\bibfnamefont {S.}~\bibnamefont
  {Ghigna}}, \bibinfo {author} {\bibfnamefont {B.}~\bibnamefont {Moore}},
  \bibinfo {author} {\bibfnamefont {F.}~\bibnamefont {Governato}}, \bibinfo
  {author} {\bibfnamefont {G.}~\bibnamefont {Lake}}, \bibinfo {author}
  {\bibfnamefont {T.~R.}\ \bibnamefont {Quinn}},\ and\ \bibinfo {author}
  {\bibfnamefont {J.}~\bibnamefont {Stadel}},\ }\bibfield  {title} {\bibinfo
  {title} {{Dark matter halos within clusters}},\ }\href
  {https://doi.org/10.1046/j.1365-8711.1998.01918.x} {\bibfield  {journal}
  {\bibinfo  {journal} {Mon. Not. Roy. Astron. Soc.}\ }\textbf {\bibinfo
  {volume} {300}},\ \bibinfo {pages} {146} (\bibinfo {year} {1998})},\ \Eprint
  {https://arxiv.org/abs/astro-ph/9801192} {arXiv:astro-ph/9801192}
  \BibitemShut {NoStop}%
\bibitem [{\citenamefont {Munoz}\ \emph {et~al.}(2008)\citenamefont {Munoz},
  \citenamefont {Majewski},\ and\ \citenamefont {Johnston}}]{Munoz:2007jn}%
  \BibitemOpen
  \bibfield  {author} {\bibinfo {author} {\bibfnamefont {R.~R.}\ \bibnamefont
  {Munoz}}, \bibinfo {author} {\bibfnamefont {S.~R.}\ \bibnamefont
  {Majewski}},\ and\ \bibinfo {author} {\bibfnamefont {K.~V.}\ \bibnamefont
  {Johnston}},\ }\bibfield  {title} {\bibinfo {title} {{Modeling The Structure
  And Dynamics of Dwarf Spheroidal Galaxies with Dark Matter And Tides}},\
  }\href {https://doi.org/10.1086/587125} {\bibfield  {journal} {\bibinfo
  {journal} {Astrophys. J.}\ }\textbf {\bibinfo {volume} {679}},\ \bibinfo
  {pages} {346} (\bibinfo {year} {2008})},\ \Eprint
  {https://arxiv.org/abs/0712.4312} {arXiv:0712.4312 [astro-ph]} \BibitemShut
  {NoStop}%
\bibitem [{\citenamefont {Wang}\ \emph {et~al.}(2017)\citenamefont {Wang},
  \citenamefont {Fattahi}, \citenamefont {Cooper}, \citenamefont {Sawala},
  \citenamefont {Strigari}, \citenamefont {Frenk}, \citenamefont {Navarro},
  \citenamefont {Oman},\ and\ \citenamefont {Schaller}}]{Wang:2016qol}%
  \BibitemOpen
  \bibfield  {author} {\bibinfo {author} {\bibfnamefont {M.~Y.}\ \bibnamefont
  {Wang}}, \bibinfo {author} {\bibfnamefont {A.}~\bibnamefont {Fattahi}},
  \bibinfo {author} {\bibfnamefont {A.~P.}\ \bibnamefont {Cooper}}, \bibinfo
  {author} {\bibfnamefont {T.}~\bibnamefont {Sawala}}, \bibinfo {author}
  {\bibfnamefont {L.~E.}\ \bibnamefont {Strigari}}, \bibinfo {author}
  {\bibfnamefont {C.~S.}\ \bibnamefont {Frenk}}, \bibinfo {author}
  {\bibfnamefont {J.~F.}\ \bibnamefont {Navarro}}, \bibinfo {author}
  {\bibfnamefont {K.}~\bibnamefont {Oman}},\ and\ \bibinfo {author}
  {\bibfnamefont {M.}~\bibnamefont {Schaller}},\ }\bibfield  {title} {\bibinfo
  {title} {{Tidal features of classical Milky Way satellites in a $\Lambda$
  cold dark matter universe}},\ }\href {https://doi.org/10.1093/mnras/stx742}
  {\bibfield  {journal} {\bibinfo  {journal} {Mon. Not. Roy. Astron. Soc.}\
  }\textbf {\bibinfo {volume} {468}},\ \bibinfo {pages} {4887} (\bibinfo {year}
  {2017})},\ \Eprint {https://arxiv.org/abs/1611.00778} {arXiv:1611.00778
  [astro-ph.GA]} \BibitemShut {NoStop}%
\bibitem [{\citenamefont {Baxter}\ \emph {et~al.}(2022)\citenamefont {Baxter},
  \citenamefont {Kumar}, \citenamefont {Paul},\ and\ \citenamefont
  {Runburg}}]{Baxter:2022dpn}%
  \BibitemOpen
  \bibfield  {author} {\bibinfo {author} {\bibfnamefont {E.~J.}\ \bibnamefont
  {Baxter}}, \bibinfo {author} {\bibfnamefont {J.}~\bibnamefont {Kumar}},
  \bibinfo {author} {\bibfnamefont {A.~D.}\ \bibnamefont {Paul}},\ and\
  \bibinfo {author} {\bibfnamefont {J.}~\bibnamefont {Runburg}},\ }\bibfield
  {title} {\bibinfo {title} {{Searching for velocity-dependent dark matter
  annihilation signals from extragalactic halos}},\ }\href
  {https://doi.org/10.1088/1475-7516/2022/09/026} {\bibfield  {journal}
  {\bibinfo  {journal} {JCAP}\ }\textbf {\bibinfo {volume} {09}},\ \bibinfo
  {pages} {026}},\ \Eprint {https://arxiv.org/abs/2205.02386} {arXiv:2205.02386
  [astro-ph.CO]} \BibitemShut {NoStop}%
\end{thebibliography}%

\end{document}